\documentstyle[12pt,epsfig]{article} 
\begin{document}
\title{Gravitomagnetism in Metric Theories: Analysis of Earth Satellites Results, and its Coupling with Spin}
\author{ A. Camacho
\thanks{email: acamacho@nuclear.inin.mx} \\
Department of Physics, \\
Instituto Nacional de Investigaciones Nucleares\\
Apartado Postal 18--1027, M\'exico, D. F., M\'exico.}
\date{}
\maketitle

\begin{abstract}
Employing the PPN formalism the gravitomagnetic field in different me\-tric theories is considered in the analysis of the LAGEOS results. It will be shown that there are several models that predict exactly the same effect that general relativity comprises. In other words, these Earth satellites results can be taken as expe\-rimental evidence that the orbital angular momentum of a body does indeed generate space--time geo\-metry, neverwithstanding they do not endow general relativity with an outstanding status among metric theories.
Additionally the coupling spin--gravitomagnetic field is analyzed with the introduction of the Rabi transitions that this field produces on a quantum system with spin $1/2$. Afterwards, a continuous measurement of the energy of this system is introduced, and the consequences upon the co\-rresponding probabilities of the involved gravi\-tomagnetic field will be obtained. Finally, it will be proved that these proposals allows us, not only to confront against future experiments the usual assumption of the coupling spin--gravotimagnetism, but also to measure some PPN parameters and to obtain functional dependences among them.
\end{abstract}
\newpage
\section{Introduction}
\bigskip

The gravitomagnetic field [1] is one of the most important predictions of general relativity (GR), which has no Newtonian counterpart, and that emerges as a consequence of mass--energy currents. Though this field has already been detected [2] it is also important to mention that this fact does not imply that Eintein's theory of gravity is {\it the correct one}. One of the reasons for this last claim stems from the fact that there are several metric theoris of gravity, this will be shown below, that predict exactly the same gravitomagnetic field as GR does, for instance, Rosen's bimetric theory of gravity [3, 4]. In other words, restricting ourselves to Ciufolini's measurement outputs we may not conclude if the perturbation of the orbit of the LAGEOS and LAGEOS II satellites was caused, for instance, by GR or by Rosen's theory. 

Clearly, at this point it may be argued that there are other experiments that discard Rosen's theory as a viable idea, nevertheless, this last argumentation can not be applied to the case of Brans--Dicke (BD) model [5], in which a free parameter can be used to reproduce the aforementioned results.  

An additional point concerns the fact that the gravitomagnetic field has been detected using classical systems, namely its effects upon the orbit of the involved satelli\-tes, but the possible consequences on quantum systems, particularly the coupling spin--gravitomagnetic field, has always been forgotten, i.e., usually it is assumed that the coupling orbital angular momentum--gravitomagnetism can be extended to explain the coupling spin--gravitomagnetic field [6]. Neverwithstanding this assumption has not only to be fathomed better, but also it must be subject to experimental scrutiny [7]. 

In this work the gravitomagnetic field of any metric theory will be used to analyze the Earth satellite results [2]. It will be shown that several theories predict the same field, GR included. Hence LAGEOS and LAGEOS II results can be taken as a proof that the orbital angular momentum of a body does indeed generate space--time geo\-metry, neverwithstanding these results do not endow general relativity with an outstanding relevance among metric theories. At most it can be claimed that GR is a viable model.

Additionally we introduce two experimental proposals that could lead to the detection 
of the coupling between intrinsic spin and the gravitomagnetic field. In order to do this, we will analyze, using the PPN formalism [4], the role that the gravitomagnetic field of the Earth, in the context of any metric theory of gravity, could have on a quantum system with spin $1/2$. In particular we will deduce, with the additional introduction of an electromagnetic field, a Rabi formula [8], which depends on the coupling between the spin of the quantum system and the gravitomagnetic field of the Earth. Afterwards, the continuous measurement of the e\-ner\-gy of the spin $1/2$ system is considered [9]. This last part of the work is an extension of a previous result [10], where only GR was considered. It will be shown that these proposals could allows us, not only to confront against future experiments the usual assumption of the coupling spin--gravitomagnetism, but also to bound some PPN parameters and to obtain functional dependences among them.
\bigskip
\bigskip

\section{Gravitomagnetic Field and PPN Formalism}
\bigskip

Let us consider a rotating uncharged, idealized spherical body with mass $M$ and angular momentum $\vec {J}$. In the weak field and slow motion limit the gravitomagnetic field may be written, using the PPN parameters $\Delta_1$ and $\Delta_2$ [11], as

{\setlength\arraycolsep{2pt}\begin{eqnarray}
\vec {H} = \Bigl({7\Delta_1 + \Delta_2\over 4}\Bigr){G\over c^2}{\vec {J} - 3(\vec {J}\cdot\hat {x})\hat {x}\over |\vec {x}|^3}.
\end{eqnarray}}
\bigskip

The case of GR implies ${7\Delta_1 + \Delta_2\over 4} = 2$, while BD appears if ${7\Delta_1 + \Delta_2\over 4} = {12 + 8\omega\over 8 + 4\omega}$. A quite interesting point emerges when we consider the situation of Ni's theory [12], where ${7\Delta_1 + \Delta_2\over 4} = 0$, i.e., there is no gravitomagnetic field.  This example shows that not every metric theory has a non--vanishing gravitomagnetic field, in this sense Ni's model is closer to the Newtonian situation. 

If we now resort to other metric theories of gravity, for instance, Rosen's [3], or Rastall's model [13], we may readily see that ${7\Delta_1 + \Delta_2\over 4} = 2$, in other words, we found three different gravitation theories that have exactly the same gravitomagnetic field. This should be no suprise at all, indeed, it is a known fact that, for instance, Rosen's theory has identical PPN parameters as those of GR, with only one exception [4].

These last cases imply that Ciufolini's measurement readouts [2] can only be taken as an indirect proof of the validity of GR, i.e.,  there are at least two additional theories of gravity that predict exactly the same gravitomagnetic field as GR does. Directly they only prove that local inertial frames of reference are influenced and dragged by mass--energy currents, and that this effect is very close to the predictions of some metric theories, GR included. Clearly these experimental outputs do discard at least one metric theory, namely Ni's model, where  ${7\Delta_1 + \Delta_2\over 4} = 0$ [4, 12].

Considering the experimental resolution [2] we may assert that the aforementioned experiment does not even discard BD, as a matter of fact if we consider that Ciufolini's  results differ from the general--relativistic predictions by a $20\%$ percent, then we may deduce that in connection with BD this means that the free parameter of this theory ($\omega$) satisfies the condition $\omega \in [-3.6, 3.0]$. In other words, not only GR, Rosen, or Rastall´s theories match with the experiment, it is also possible to find and interval for $\omega$ such that BD reproduces the measurement outputs.

Some interesting conclusions may be drawn if we consider Lee's theory [4, 14]. If we analyze this model in the context of the experimental results [2], then we may impose bounds upon an expression that involves three of its PPN parameters ($e$, $c_0$, and $c_1$) 

{\setlength\arraycolsep{2pt}\begin{eqnarray}
{e\over 4\sqrt{c_0c_1}} \in [0.9, 1.3],
\end{eqnarray}}
\bigskip
 
\noindent which means that $e$ is a function of $c_0c_1$. In other words, the arbitrary parameter $e$ is, through the gravitomagnetic effect, given as a function of the cosmological variables $c_0$ and $c_1$. Additionally from this functional dependence the two remaining arbitrary  parameters of this model [4, 14], $\kappa_1$ and $\kappa_2$, are also fixed as a function of the cosmological variables.

\bigskip
\bigskip

\section{Coupling Spin--Gravitomagnetism in Metric Theories}
\bigskip

Let us consider a spin $1/2$ system immersed in the gravitomagnetic field given by expression (1). We will assume that the expression that describes the precession of orbital angular 
momentum can be also used to the description of the dynamics in the case of intrinsic spin. In the context of GR this is a usual assumption [6], though if we adopt a more critical position this fact has to be put under experimental scrutiny, moreover it has to be underlined that up to now there is a lack of experimental evidence in this direction. 

Let us now denote the angular momentum of our spherical body by $\vec {J} = J\hat {z}$, 
being $\hat {z}$ the unit vector along the direction of the angular momentum.
Our quantum particle is prepared such that $\vec {S} = S_z\hat {z}$.
For simplicity we will also assume that our system carries vanishing small velocity and 
acceleration, and that it is located on the $z$--axis, with coordinate $Z$.

Following this analogy between gravitomagnetism and magnetism we may now write down 
the interaction Hamiltonian (acting in the two--dimensional spin space of our spin $1/2$ system)
which gives the coupling between $\vec {H}$ and the spin, $\vec {S}$, of our particle.
\bigskip

{\setlength\arraycolsep{2pt}\begin{eqnarray}
H = - \vec {S}\cdot\vec {H}.
\end{eqnarray}}
\bigskip

With (1) we may rephrase this Hamiltonian as

{\setlength\arraycolsep{2pt}\begin{eqnarray}
H = \Bigl({7\Delta_1 + \Delta_2\over 4}\Bigr){GJ\hbar\over c^2Z^3}\Bigl[|+><+| - |-><-|\Bigr].
\end{eqnarray}}
\bigskip
 
Here $|+>$ and $|->$ represent the eigenkets of $S_z$. 
Clearly, the introduction of the gravitomagnetic field renders two energy states, 
i.e., it breaks the existing symmetry in the Hilbert spin--space.

{\setlength\arraycolsep{2pt}\begin{eqnarray}
E_{(+)} = \Bigl({7\Delta_1 + \Delta_2\over 4}\Bigr){GJ\hbar\over c^2Z^3},
\end{eqnarray}}
\bigskip

{\setlength\arraycolsep{2pt}\begin{eqnarray}
E_{(-)} = -\Bigl({7\Delta_1 + \Delta_2\over 4}\Bigr){GJ\hbar\over c^2Z^3},
\end{eqnarray}}
\bigskip

\noindent where $E_{(+)}$ ($E_{(-)}$) is the energy of the spin state $+\hbar/2$ ($-\hbar/2$). 
Let us now define the frequency

{\setlength\arraycolsep{2pt}\begin{eqnarray}
\Omega = (E_{(+)} - E_{(-)})/\hbar = 2\Bigl({7\Delta_1 + \Delta_2\over 4}\Bigr){GJ\over c^2Z^3}.
\end{eqnarray}}
\bigskip

But this analogy between gravitomagnetism and magnetism allows us to consider the 
emergence of Rabi transitions [8] (with the introduction of a rotating magnetic field), in which the transition probabilities will 
depend upon the coupling between the intrinsic spin of our quantum system and the gravitomagnetic field 
of $M$. Thus, in principle, gravitomagnetism could be detected by means of Rabi transitions. 

In order to do this let us now introduce a rotating magnetic field, which, at the point where the particle is 
located, has the following form

{\setlength\arraycolsep{2pt}\begin{eqnarray}
\vec {b} = b\Bigl[cos(wt)\hat {x} + sin(wt)\hat {y}\Bigr],
\end{eqnarray}}
\bigskip

\noindent where $\hat {x}$ and $\hat {y}$ are two unit vectors perpendicular to the $z$--axis, 
and $b$ is a constant magnetic field.

Looking for a solution in the form $|\alpha> = c_{(+)}(t)|+> +~c_{(-)}(t)|->$, we find, assuming that our quantum system has been initially prepared such that $c_{(-)}(0) = 1$ and 
$c_{(+)}(0)= 0$, that 
\bigskip

{\setlength\arraycolsep{2pt}\begin{eqnarray}
{|c_{(-)}(t)|^2\over |c_{(-)}(t)|^2 + |c_{(+)}(t)|^2} = 
\Bigl[ 1 + {({eb\over 2mc\Gamma})^2sin^2(\Gamma t)\over cos^2(\Gamma t) + 
{(w - \Omega)^2\over 4\Gamma^2}sin^2(\Gamma t)}\Bigr]^{-1}.
\end{eqnarray}}
\bigskip

Rabi transitions depend upon the coupling between spin and the gravitomagnetic field, $\Gamma = \sqrt{({eb\over 2mc})^2 + {(w - \Omega)^2\over 4}}$. This frequency allows us, in principle, to detect the metric theory that generates this transition, up to a class defined by those metric theories that have the same value of ${7\Delta_1 + \Delta_2\over 4}$. 

Let us now measure the energy of our spin $1/2$ system, such that this experiment lasts a time $T$ and $\Delta E$ denotes the re\-solution of the experimental device. From previous work [10] 
we expect the inhibition of the evolution of our system, this is the so called quantum Zeno effect. Proceeding in a similar way as in [10] we find that the probability of having as a measurement output the state $|->$ is (assuming that ${(E_{(+)} - E_{(-)})^4\over 4T^2\Delta E^4}> \gamma^2/\hbar^2$) is
 
{\setlength\arraycolsep{2pt}\begin{eqnarray}
P_{(-)}(t) = 
\Bigl\{1 + {\sinh^2({\gamma\over\hbar}\tilde{\Omega}t)\over 
\tilde{\Omega}^2
[\cosh({\gamma\over\hbar}\tilde{\Omega}t) + {\hbar(E_{(+)} - E_{(-)})^2\over 2T\gamma\tilde{\Omega}\Delta E^2}
\sinh({\gamma\over\hbar}\tilde{\Omega}t)]^2}\Bigr\}^{-1},
\end{eqnarray}}
\bigskip

\noindent where $\tilde{\Omega} = \sqrt{{\hbar^2(E_{(+)} - E_{(-)})^4\over 4T^2\gamma^2\Delta E^4} - 1}$, $\gamma = -{eb\hbar\over 2mc}$.

In order to compare the predictions coming from different metric theories, let us now denote with $\tilde{\Omega}^{(r)}$ and $\Omega^{(r)}$ the corresponding frequencies that appear when 
${7\Delta_1 + \Delta_2\over 4} =2$, a condition that comprises GR. Then the last probability becomes ($\alpha = {\vert 7\Delta_1 + \Delta_2\vert\over 8}$)

{\setlength\arraycolsep{2pt}\begin{eqnarray}
P_{(-)}(t) = 
\Bigl\{1 + \Bigl[\alpha^2\tilde{\Omega}^{(r)}\coth(\alpha^2{\gamma\over\hbar}\tilde{\Omega}^{(r)}t) + \alpha{\hbar^3(\Omega^{(r)})^2\over 4T\gamma\Delta E^2}\Bigr]^{-2}\Bigr\}^{-1}. 
\end{eqnarray}}
\bigskip

Expression (11) gives us the possibility of imposing bounds on some PPN parameters. For instance, if we take Lee's gravitational theory [14] (where ${\vert 7\Delta_1 + \Delta_2\vert\over 8} = {e\over 4\sqrt{c_0c_1}}$), then we may obtain ${e\over 4\sqrt{c_0c_1}}$ as a function of $P_{(-)}(t)$ and of the remaining experimental variables. Hence we would obtain, for instance, $e$ as a function of $\sqrt{c_0c_1}$. This renders a possibility of testing the validity of the assumption concerning the interaction betwen spin and gravitomagnetism. Indeed, in this context we have a functional dependence between $e$ and $\sqrt{c_0c_1}$, and this result could be confronted with expression (2)), which is a consequence of an experiment that involves orbital angular momentum, and not spin.

It is readily seen that though here we have employed Lee's model the aforementioned argument may be used to analyze the validity of this ubiquitous assumption in the context of other metric theories of gravity, including GR. For instance, in the case of the so called BSLL bimetric theory [15], two arbitray parameters, $a$ and $K$, and two cosmological variables, $\omega_0$ and $\omega_1$, would be involved in the corresponding functional dependences.
In other words, Ciufolini's experiment, and the results that would come from (11), would allow us to refute or validate, upon experimental grounds, this Ansatz. This would be done finding the bounds upon the corresponding PPN parameters, that [2] imposes, and then check them with those bounds appearing as a consequence of (11).
\bigskip
\bigskip

\section{Conclusions}
\bigskip

In this work the gravitomagnetic field of any metric theory has been used to analyze the Earth satellite results. It was shown that there are several theories which have exactly the same kind of prediction that appears in connection with GR. Hence LAGEOS results are only an indirect proof of the validity of GR. Nevertheless they allow us, not only to discard some theories, for instance, Ni's model, but also to find some bounds and functional dependences among PPN parameters.

In this context it is noteworthy to comment that the bounds found for the BD case ($\omega \in [-3.6, 3.0]$) have an interesting connection with the existence of black holes that are absent in GR [16], where this possibility appears if $\omega \in [-5/2, -3/2)$. This last statement may be rephrased claiming that the LAGEOS results contain the possibility of having black holes which would not appear in GR.

Additionally we introduced two experimental proposals that could allow us to confront, against future experiments, the usual assumption concerning the interaction spin--gravitomagnetism. It was shown that our quantum expression depends critically upon the PPN parameters of the involved theory, and therefore they could be employed as tool to check, experimentally, the validity of this Ansatz, where Ciufolini's results would be used as a confronting framework for the results obtained at quantum level.

The confirmation of the Zeno quantum effect [17] sets a sound background for the present proposal, though several experimental issues need to be addressed yet, in order to say something about its feasibility, for instance, how to screen the quantum system from external perturbations, as the magnetic field of the Earth.
\bigskip
\bigskip

\Large{\bf Acknowledgments.}\normalsize
\bigskip

The author would like to thank A. A. Cuevas--Sosa for his 
help. This work was partially supported by CONACYT (M\'exico) Grant No. I35612--E.

\end{document}